# MULTIPLE LAYERS OF FUZZY LOGIC TO QUANTIFY VULNERABILITIES IN IoT

Mohammad Shojaeshafiei[1], Letha Etzkorn[1], and Michael Anderson[2]

[1]Department of Computer Science, The University of Alabama in Huntsville, Huntsville, USA
Ms0083@uah.edu, etzkorl@uah.edu

[2] Department of Civil and Environmental Engineering, The University of Alabama in Huntsville, Huntsville, USA
andersmd@uah.edu

## ABSTRACT

*Quantifying vulnerabilities of network systems has been a highly controversial issue in the fields of network security and IoT. Much research has been conducted on this purpose; however, these have many ambiguities and uncertainties. In this paper, we investigate the quantification of vulnerability in the Department of Transportation (DOT) as our proof of concept. We initiate the analysis of security requirements, using Security Quality Requirements Engineering (SQUARE) for security requirements elicitation. Then we apply published security standards such as NIST SP-800 and ISO 27001 to map our security factors and sub-factors. Finally, we propose our Multi-layered Fuzzy Logic (MFL) approach based on Goal question Metrics (GQM) to quantify network security and IoT (Mobile Devices) vulnerability in DOT.*

## KEYWORDS

*Computer Network, Network Security, Mobile Devices, Fuzzy Logic, Vulnerability, Cyber Security*

## 1. INTRODUCTION

Today we see drastic development and improvement of Information and Communication Technology (ICT). It is inevitable that many changes in computer networks result in security complications. Thus, when we analyze an organization's network system we are required to consider its potential vulnerabilities for any unexpected attack or information leakage. many types of methodologies and procedures have been proposed using proper measurement approaches to tackle such vulnerabilities in an organization's network system [1][2][3]. Mainly, these methodologies are divided into two categories: Qualitative and quantitative [4]. In the past [5] Multi-layered Fuzzy Logic has been used to quantify previously qualitative concepts. In this paper, we propose a Multi-layered Fuzzy Logic (MFL) approach to quantify potential vulnerabilities of the Department of Transportation (DOT). It comprises the computer network's security and Mobile Devices as regards to IoT of DOT. For that, we have considered all aspects of DOT's computer networks and Mobile Devices for security analysis. With a thorough security analysis to measure vulnerabilities, we have listed all security aspects of computer network and Mobile Devices as the security factors in a top-down manner from major to minor. Afterward, we evaluated all security factors to check if they overlap (in order to remove redundancy). Finally, we accepted Availability, Integrity, Accuracy, and Confidentiality for computer network vulnerability analysis, and Enterprise Mobile Management (EMM), User Access Control(UAC), and Encryption for Mobile Devices for consideration.

We apply the Goal Question Metrics (GQM) [6] approach to provide the required input for the MFL approach. The input for MFL will be in the form of qualitative processes that are derived from the computer network and security expert evaluation in DOT based on standard security questions mapped to NIST SP800-53 and ISO 27001 standards. The entire procedure is addressed in section 3.

The rest of this paper is organized as follows: Section 2 presents the background of the work to address the Fuzzy Logic in related work, computer network vulnerability quantification and IoT's vulnerability measurement. Section 3 provides a description of our methodology in GQM, security standards, factors and sub-factors of network security and Mobile Devices. Section 4 addresses Multi-layered Fuzzy Logic (MFL) implementation.

## 2. BACKGROUND

### 2.1. Fuzzy Logic and Related Work

Vulnerability measurement processes almost always result in a high degree of uncertainty. Because whenever we discuss the security of a network, we describe it as a linguistic variable form of 'secure', or 'not secure', that causes imprecision and vagueness. Therefore, we are not able to evaluate the accuracy of evaluation based on linguistic variables. In 1965, Lotfi Zadeh [7] at the University of California at Berkley proposed Fuzzy logic (and proved it mathematically). This method says that conventional computer logic is not able to work on data manipulation when the data carries the vagueness of human linguistic propensity. One of the advantages of the Fuzzy Logic methodology for vulnerability measurement is that the implementations based on mathematical models are reliable in all aspects of security analysis based on previous analysis of indication motors [8].

It is very important to know the degree of truth in Fuzzy Logic. Membership Function (MF) [9] is a pivotal component in Fuzzy Logic to provide such a degree of truth. It defines a function that specifies the degree of "belongingness" of an input to a set. The value of MF is always limited in the interval of [0-1]. The most common forms of MF are Triangular, Trapezoidal, piecewise Linear, Gaussian, and Singleton [10]. Figure 1 depicts an example of Triangular MF for temperature in a sub-set of `cold, cool, normal, warm` and `hot`.

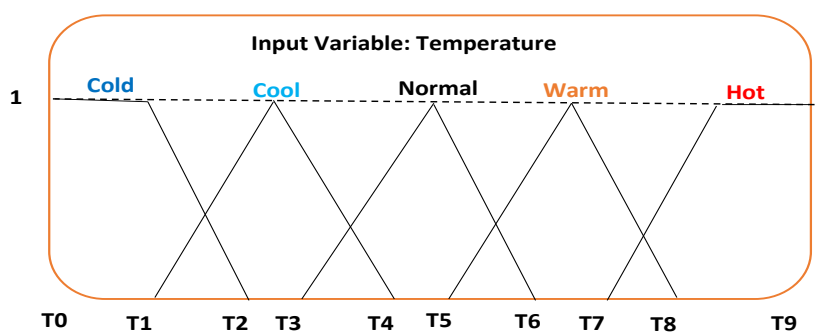

Figure 1. Triangular MF for Temperature

The triangular function is defined as follows:

" For a fuzzy set, A on the universe of discourse X is defined as $\mu_A: X \to [0,1]$, where each element of X is in an area between 0 and 1 that quantifies the grade of membership of the element in X to the fuzzy set A. It is defined by a lower bound a and an upper bound b and the value m where a<m<b " [7] (Table 1).

Table 1. Triangular Function Equation.

| $\mu_A(x)=$ | 0 | If: $x \leq a$ |
|---|---|---|
| | $(x-a)/(m-a)$ | If: $a < x \leq m$ |
| | $(b-x)/(x-m)$ | If: $m < x \leq b$ |
| | 0 | If: $x \geq b$ |

A Fuzzy Logic system is a nonlinear mapping of the input data to a scalar output data and it has four major steps as shown in Figure 2.

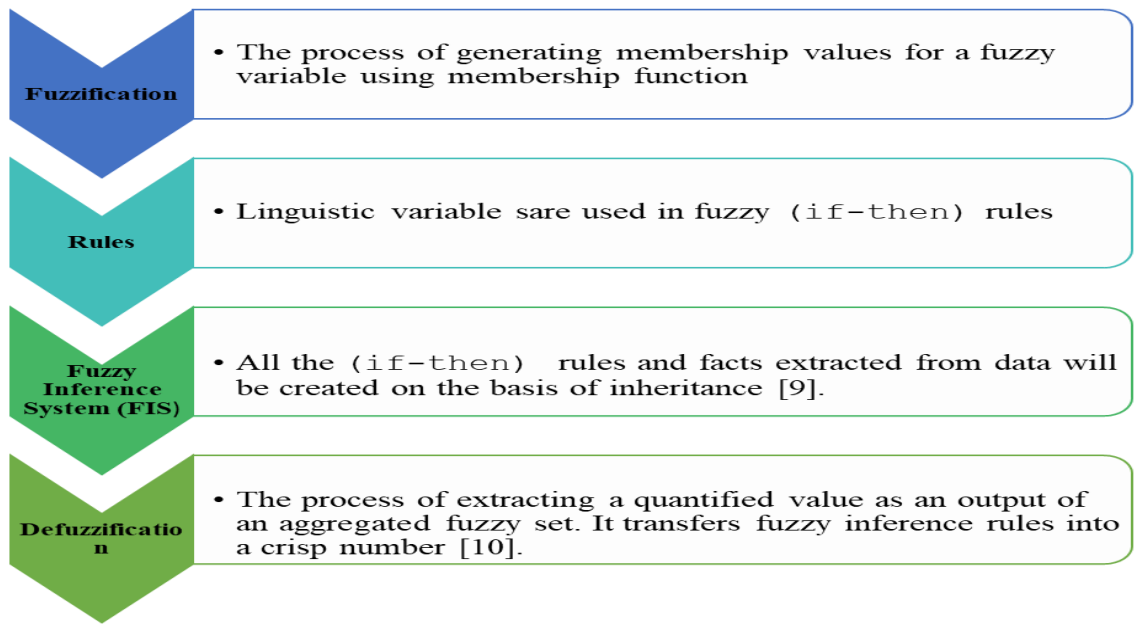

Figure 2. Fuzzy Logic Steps

Some authors have conducted research using the Mamdani model [13] of Fuzzy Inference to identify and measure potential threats and risks [14][15]. They applied this model to software to find out if the software modules are prone to attack or not. The main goal of their methodology was to eliminate training and testing phases while building up the procedure of software fault prediction using soft computing methods.

Kamongi et al. [16] proposed a method of ranking cloud system vulnerability and implemented it with vulnerability discovery on a cloud web application: the result was a list of ranked vulnerabilities associated with an attack path. They conceptualized cloud vulnerabilities as the attack paths from a pre-generated attack graph. Each path consists of information in regard to vulnerabilities of pre and post conditions that are required to be met for a successful security breach.

Anikin et al. [17] proposed a quantitative information security risk assessment in computer networks. They applied Fuzzy Logic and analytic hierarchy processes to evaluate the impact and quantify the value of a specific threat. They presented the vulnerability risk assessment based on

Vulnerability Scoring System (CVSS) [18], but with the elimination of CVSS' barriers to risk quantification. They suggested fuzzy `if-then` rules for decision support systems with discrete output that was obtained from decision making in oil production.

Kuang et al. [19] conducted research on network security situation forecast using Fuzzy logic based on the Markov model. Their methodology is based on the combination of safe behavior historical data with the level of threats in the system. They analyzed branch prediction based on the Markov model and membership degree evaluation based on the Fuzzy system of information security vulnerabilities. Finally, they provided an integrated version of security situation prediction based on Fuzzy Logic. For verification, they used data from KDD CUP99's data as training data and the DARPA2000's data as testing data to obtain vulnerability information [20][21].

## 2.2. Internet of Thing (IoT) Vulnerability Assessment

Internet of Things (IoT) is defined as the structure and combination of multiple devices that are connected to each other through the internet [22]. Interaction with and between IoT devices has changed over time. These changes can be due to either software or hardware variations. Such changes, which result in a loss of integrity, are nowadays the major subject of security concerns in IoT. Lack of integrity in IoT causes a considerable cost for service providers and users because the adversary is able to take advantage of such security holes and cause irreparable harm. Millions of IP security cameras, doorbells, etc. are vulnerable to attack or can be hijacked to work against the devices' owners. In organizations' security policies, the inherent risk of Bring Your Own Device (BYOD) made the IT departments to become more restrictive and required extensive monitoring of portable devices. Mobile Device Management (MDM) tools are also applied as security software to monitor and manage employees' mobile devices. However, with all of these strict policies and monitoring, a wide variety of vulnerabilities still threatening organizations and put them into a serious risk of the data breach. Much research has been studied in this area as a major concern of today's internet security. However, the problems in this area are not yet solved.

Williams et al. [23] conducted research on vulnerability assessment of consumer IoT devices. They used Nessus [24] for vulnerability assessment since it has the capability of scanning many devices at a time. They scanned a large number of devices and extracted the result in a set such as 'critical', 'high', 'medium', and 'low' risks. They used Nessus for vulnerability assessment of IoT in different categories including devices of the home, workplace, and cities.

Patton et al. [25] studied the vulnerable devices on the IoT and provided an evaluation of existing vulnerabilities in the IoT system. They applied Shodan, a search engine for IoT, that uses its database to maintain past scans. In the first step, they built an executed version of python scripts to utilize the Shodan API to interface with Shodan. Then they parsed device headers into MySQL database and used a password database; with a python script, they tested the password against IoT devices to capture vulnerabilities. They used the list of IP addresses collected from Shodan to scan thousands of devices to check if the default login credentials work. The result of vulnerabilities varied from %0.44 to %40.

### 2.2.1. Mobile Devices

Based on Gartner Research's forecast [26] there will be 20.4 billion IoT devices by the end of 2020 and almost 75 billion devices in 2025. With the expansion of the network systems in organizations, increasing the number of IoT devices due to the immense availability of connectivity devices is inevitable. This expectation of exponential proliferation is reasonable because enterprises make a large profit through the application of these ubiquitous mobile devices.

On the other hand, IoT (including all mobile devices) put the enterprises at a high risk of being susceptible to attack through potential vulnerabilities. One of the most important parts of the cybersecurity domain is vulnerability measurements in Mobile Devices as a portable form of IoT in organizations such as DOT. There have been many arrangements for Mobile Devices policies and restrictions including building multiple layers of protection in agencies to keep their information safe from cyber-attacks. The more the employees in agencies use mobile devices the more the probability of cyber-attacks will emerge. The support layers to bolster the security of Mobile Devices consist of providing anti-malware software, secure mobile communications (using VPNs and requiring strong encryption and authentication), control of third-party's software, performing penetration testing to check for vulnerabilities, auditing, etc. Nevertheless, a clear-cut quantification procedure to measure vulnerabilities of Mobile Devices has always been neglected, or at least, it has not been taken as seriously as possible. Thus, in this paper, we consider all aspects of Mobile Devices as part of IoT in DOT to provide a quantifiable measurement methodology for the potential vulnerabilities of a system.

## 3. METHODOLOGY

### 3.1. Security Requirements

Prior to the vulnerability measurement process of network security and Mobile Devices, we have to assure that our network design and implementation comply with standard security requirements as a pre-requisite of network design. One of the most important security flaws in organizations that leads them to be prone to cybersecurity attacks is the lack of a comprehensive and precise engineered design of security requirements [27]. It is implausible to design security requirements for a network system that guarantee no flaws, but still, organizations must take this step as a fundamental stage of attack prevention. The detection and correction phase of security requirements development costs 10 to 200 times less than the detection of flaws after system deployment in the field [27]. For security requirements elicitation there have been several proposed models such as Multilateral Security Requirements Analysis (MSRA) and Goal Base Requirements Analysis Record (GBRAM). In this paper, we apply the Security Quality Requirements Engineering (SQUARE) [28] for security requirements elicitation because it considers risk assessment and quality assurance simultaneously to maintain availability, integrity, and confidentiality of the security requirement goals. SQUARE is a process model developed at Carnegie Mellon University to elicit, categorize and prioritize security requirements for IT departments and applications.

In order to generate final deliverable prioritized security requirements in the system, SQUARE provides nine fundamental steps for security requirements elicitation. We adopted this model and extracted several major points to be highlighted as fundamental security requirements in DOT. These steps are Integrity, Physical Security, Authentication, Access Control, Availability, Audition, and Authorization. Therefore, the very first step of vulnerability measurement in DOT is following these security requirement elicitation processes to design all components of security in the system for network security and Mobile Devices (IoT).

### 3.2. Goal Question Metrics

Goal Question Metrics (GQM) is an approach for developing a model based on the goals of the project. This model was originally developed by Basili [29] at the University of Maryland to measure software quality. The metric measurement model is proposed in three steps: conceptual model (Goal), operational model (Question) and quantitative model (Metrics) [30]. One good aspect of this model's application is that there may be many goals and consequently multiple times questions for a project, but the rate of metrics does not grow at the same rate as goals or questions. These three steps are decomposed to six steps in detail for metric measurements in the software product as follows:

1. Develop project goals.
2. Generate questions that describe the goals in step 1.
3. Define the specific measures to answer the questions designed in step 2.
4. Data collection mechanisms.
5. Validate and analyze data that are obtained in step 4.
6. Data analysis to assess the conformance of the goals.

Based on the analysis and consideration of security requirements using SQUARE model for network security and Mobile Devices in DOT we design questions accordingly for each security factor in a top-down manner to reach the desired goal (vulnerability quantification). The main reason for using GQM in our research is that it is a de facto standard for quality metrics in software engineering and is therefore widely accepted and respected [29]. Our use of GQM is analogous to how it has been previously used in software metrics. Based on the security requirements of DOT we designed the goals (numerical vulnerability value of each security factor) that are traceable to a set of quantifiable questions [31]. In our previous research [32], we introduced a hierarchy model to identify the security requirements of organizations using GQM to be a primary work of vulnerability measurement that enable us to trace from security requirements to security metrics.

### 3.3. Security Standards

After identifying the security requirements and determining security goals and questions based on GQM to measure vulnerability, we map every possible security question derived from essential factors of network security and Mobile Devices to the currently published security standards from NIST SP800-53 and ISO 27001. Passing through this process is to make sure that we are following all the available security standards.

ISO/IEC 27001:2013 [33] is an international standard to specify an information security management system (ISMS). This standard helps organizations focus on three key aspects of information: confidentiality, integrity, and availability. It examines the organization's risk, threat and potential vulnerabilities, provides all information security controls in order to reinforce strategies for risk avoidance, and finally, it makes sure that all security management processes and security controls meet the organizations' security needs.

Another purpose of this standard is to verify the proper selection of proportionate security controls to provide an acceptable amount of protection for assets. In sum, its main features are information security policies, communication, and operational management, access control, information system acquisition, organization of information security, asset management, business continuity management, human resources security and physical security [34].

NIST SP800-53 [35] is a publication of the National Institute of Standards and Technology (US) that specifies a set consisting of 198 security controls. Controls are categorized into three main groups: technical, operational and management. These three groups can characterize several subgroups such as access control, awareness and training, audit and accountability, security assessment and authorization, configuration management, contingency planning, identification and authorization, system and information integrity, etc.

### 3.4. Security Factors and Sub-factors in DOT

While analyzing the main components of network security and Mobile Devices (as part of IoT) in DOT, we designed each security component as security factors and sub-factors at each level. Table 2 and Table 3 describe each security factor and sub-factor based on the published security standards of NIST SP800-53 and ISO 27001.

Table 2. Network Security Factors and Sub-factors in DOT.

| | | | |
|---|---|---|---|
| Network Security | Availability | Redundancy | N/A |
| | | Monitoring & Alerting | |
| | | Backup | |
| | | Load Balancing( proxy) | |
| | | Disaster Readiness | |
| | | Disaster Recovery Plan | |
| | | Quality Testing | |
| | | Wireless Security | |
| | Accuracy | Data Accuracy | Annual Review |
| | | | VAM(Vulnerability Assessment & Management) |
| | | | Data labeling |
| | Integrity | Data security | Encrypted Communication |
| | | | Supply Chain |
| | | Data Quality and Integrity | N/A |
| | | Software Security | Software Update |
| | | | Outsource Development |
| | | | Threat Detection Update |
| | | Quality Testing | N/A |
| | | Wireless Security | N/A |
| | Confidentiality | Authorization and Identification | Username and Password |
| | | | Program Access |
| | | Authentication | Data Classification, |
| | | | Two-factor Authentication |
| | | | Use Geographic Location as an Authentication Factor |
| | | User Access Control (UAC) | N/A |
| | | Encryption | N/A |
| | | Quality Testing | N/A |

Table 3. Mobile Devices' Security Factors and Sub-factors in DOT.

| | | | | |
|---|---|---|---|---|
| Mobile Devices | Enterprise Mobility Management (EMM) | Mobile Device Management System ( MDM) | Lost Devices | Security Questions or Challanges |
| | | | | Lost or Stolen Reports |
| | | | Inventory of Mobile Devices | |
| | | | Automatic Lockout Screen | |
| | | | Jailbreaking Prevention | |
| | | Mobile Application Management System (MAM) | Operating System Management Program | |

| | | Remote Data Wipe | MDM |
|---|---|---|---|
| | User Access Control (UAC) | | EMM |
| | | Username and Password | Credential Prompt |
| | | | Log Viewer |
| | | External Memory Policy | N/A |
| | | Bring Your Own Device (BYOD) | N/A |
| | | Monitoring Policy | N/A |
| | Encryption | Data and Device Encryption | Encryption Algorithms |

Since our goal in this research is to quantify vulnerabilities in DOT from the perspective of network security and IOT, we designed relevant security questions to map all security factors and sub-factors mentioned in Table 2 and Table 3 based on the GQM model. The questionnaire is designed in a top-down manner to assign one or more question(s) for vulnerability measurement depending on the type of questions. The questionnaire will be answered by the computer network and security experts of DOT. A few numbers of questions are chosen from the questionnaire and are listed in Table 4.

Table 4. Security Questions derived from GQM mapped to Security Standards

| Security Factor | Question |
|---|---|
| Availability | Does DOT make sure security mechanisms and redundancies are implemented to protect equipment from utility service outages (e.g., power failures, network disruptions, etc.) |
| Integrity | How often does DOT ensure that data does not migrate beyond a defined geographical residency? |
| Accuracy | How often does your organization consider annual review including third party providers upon which their information supply chain depends? |
| Confidentiality | Does your agency require two-factor authentication for remote access? (e.g. token is used in addition to a username, and password). |
| Enterprise Mobility Management | How often does your agency require/remind employees to report their mobile devices' lost or stolen? |
| User Access Control (UAC) | Does your organization require and enforce via technical controls an automatic lockout screen for mobile devices or any company-owned devices? |
| Encryption | How often does your organization's mobile device policy require the use of encryption for either the entire device or for data identified as sensitive enforceable through technology controls for all mobile devices? |

## 3.5. Multi-Layered Fuzzy Logic

The main reason for using Fuzzy Logic is that it is one of the most reliable mathematical tools to model problems that have the most inaccuracy and uncertainty [7]. Another crucial reason for using Fuzzy Logic in this research is that we are dealing with linguistic variables to qualitatively determine the value of the security as 'good' or 'bad'. As we described the Triangular Fuzzy model MF in section 2, we will apply it for vulnerability quantification because it provides a simple Fuzzy Inference System (FIS) that correlates the vulnerability attributes quantitatively in fuzzification processes. We use the most common properties of fuzzy logic for $A_{fuzzy}$ and $B_{fuzzy}$ as follows:

$$\mu_{A \cup B}(x) = \min[\mu_A(x), \mu_B(x)] \mid x \in X \quad , \quad \mu_{A \cap B}(x) = \max[\mu_A(x), \mu_B(x)] \mid x \in X$$

In the Defuzzification process, this model generates a crisp number derived from the fuzzy set. There are several Defuzzification methods such as Center of Area (COA), Bisector of Area (BOA), largest of Maximum (LOM), Mean of Maximum (MOM), Smallest of Maximum, etc [36]. We use the COA method [37] since this method considers, determines the center of the area of fuzzy set and returns the corresponding crisp value of that [38][12].

To measure the vulnerability of the network system and Mobile Devices in DOT, we follow a few important steps from the analysis part in the cybersecurity domain to the final step of measurement. Figure 3 depicts the order of the processes.

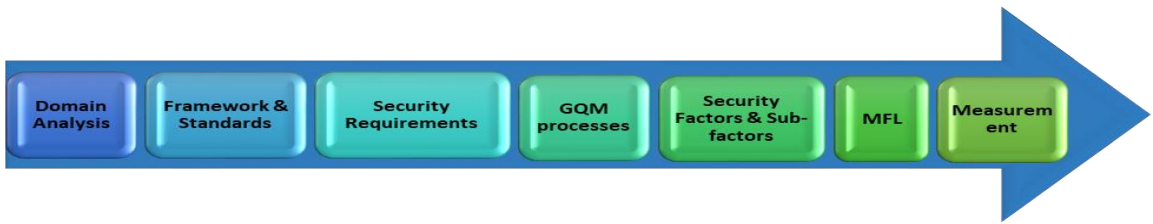

Figure 3. vulnerability Measurement Steps in DOT

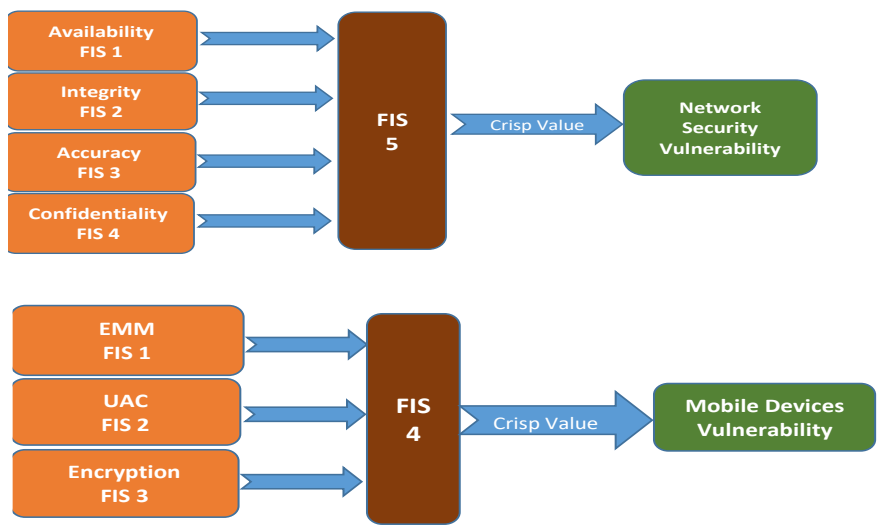

Figure 4. MFL for Network Security and Mobile Devices (IoT) in DOT

As it shows in Figure 4 we define a Fuzzy Inference System (FIS) process separately for each security factor in a backward flow from sub-score (leaves) to the main factors (nodes). In the next phase, the results of the first-step-FIS are being processed cumulatively in the second-step-FIS. The result is a crisp value to measure vulnerabilities of the factors based on the FIS rules that are defined for linguistic variables. The main steps to determine the parameters of FIS are the MF calculation mechanism of linguistic values to maintain them in a database for both antecedent (`if`) and consequent (`then`), plus the Fuzzy reasoning mechanism from the number of used Fuzzy rules [39].

When security experts in the DOT answer the questionnaire for each sub-factor, the answers will be categorized to appropriate security Groups based on the role of each sub-factor in network security, and Mobile Devices. The reason for this classification is that the answers are in the format of fuzzy subsets such as 'very low', 'low', 'medium', 'high' and 'very high'. The weights from 0 to 10 are assigned to each sub-factor in the questionnaire to determine what fuzzy subset each sub-factor belongs to. In this case, the factor Enterprise Mobility Management (EMM) is the result of the combination of all sub-factors in Group 1 (Mobile Device Management System (MDM)) and Group 2 (Mobile Application Management System (MAM)). To achieve MDM and MAM we apply FIS accordingly based on their own sub-groups derived from Table 3. As shown in Figure 5 the interior FIS layers of MDM are generated from security questions, lost or stolen reports, jailbreaking prevention, automatic lockout and inventory of mobile devices. After obtaining all required crisp values from MFL, we measure the final value of vulnerability from the aggregation of the previous values.

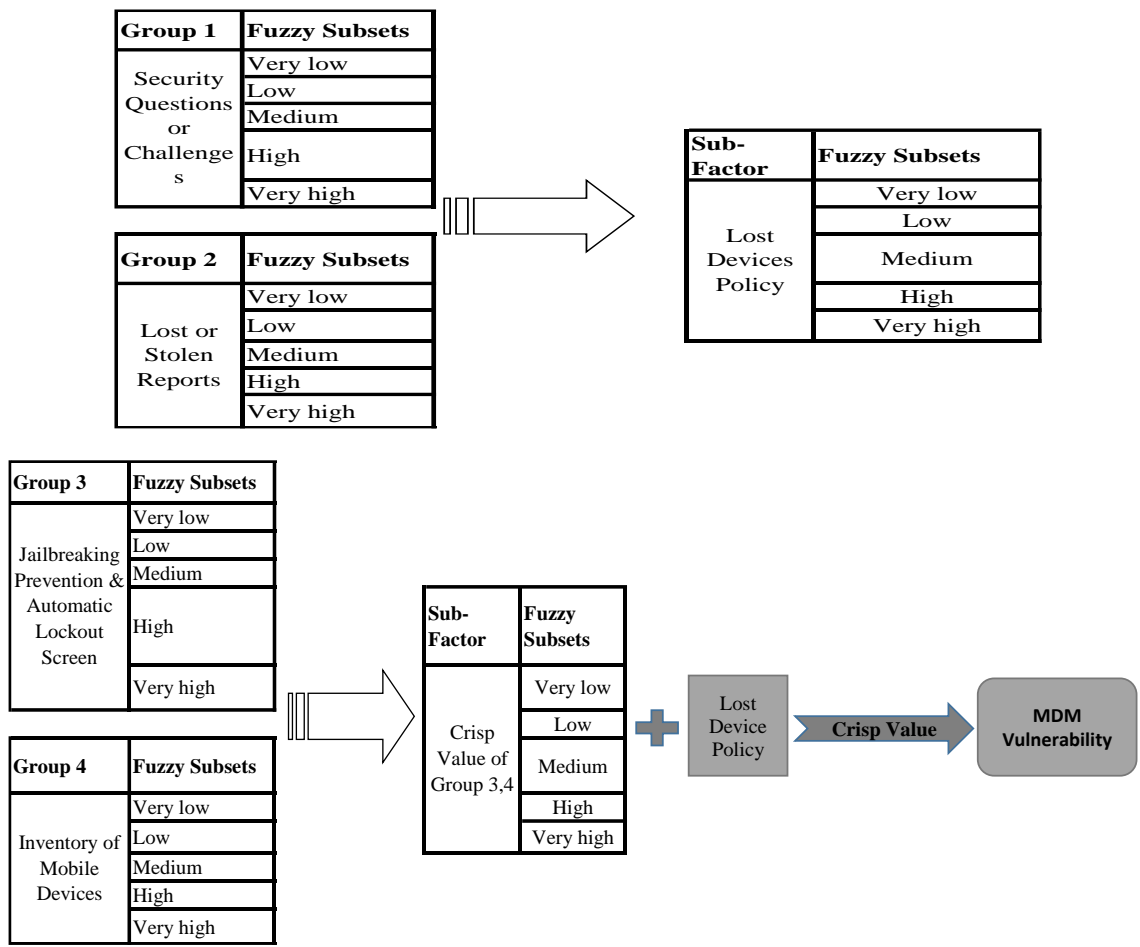

Figure 5. Fuzzy Subset implementation for EMM based on interior FIS.

In order to convert the expert answers of the questions from the form of linguistic values to the form of Fuzzy subset, we need to define the concepts of the values 'very low', 'low', 'medium', 'high' and 'very high' for each question. Since the implementation and the full discussion of all components of network security and Mobile Devices are beyond the scope of this paper we show only the implementation of Mobile Devices (IoT) in DOT. Table 5 addresses these definitions based on the answers to each question in Mobile Devices' factors and sub-factors.

Table 5. Description of Weighted Questions to convert to Fuzzy Subsets

| Sub-factors | | Description in Fuzzy sets |
|---|---|---|
| Security Questions | Very low | DOT never asks security questions or challenges to their employees to prevent unwanted access to Mobile Devices. |
| | Low | DOT sometimes asks security questions to their employees to prevent unwanted access to Mobile Devices. |
| | Medium | DOT regularly provides mechanisms to prevent unwanted access to Mobile Devices. |
| | High | DOT usually provides mechanisms to prevent unwanted access to Mobile Devices. |
| | Very High | DOT always asks extra security questions and challenges to their employees coupled with providing mechanisms to prevent unwanted access to Mobile Devices. |
| Lost or Stolen Reports | Very Low | DOT never requires their employees to report if their Mobile Devices are lost or stolen. |
| | Low | DOT rarely requires their employees to report if their Mobile Devices are lost or stolen. |
| | Medium | DOT regularly requires their employees to report if their Mobile Devices are lost or stolen. |
| | High | DOT usually requires their employees to report if their Mobile Devices are lost or stolen, if so, they usually have a security plan for data protection of such devices. |
| | Very High | DOT always requires their employees to report if their Mobile Devices are lost or stolen, if so, they always have a strong security plan for data protection of such devices. |
| Inventory of Mobile Devices | Very Low | DOT never maintains an inventory of all Mobile Devices. |
| | Low | DOT sometimes maintains an inventory of all Mobile Devices. |
| | Medium | DOT regularly maintains an inventory of all Mobile Devices. |
| | High | DOT usually maintains an inventory of all Mobile Devices storing and accessing company data which includes the status of OS, patch level, lost, decommissioned and device assignee. |
| | Very High | DOT always maintains an inventory of all Mobile Devices storing and accessing company data which includes the status of OS, patch level, lost, decommissioned and device assignee. |
| Automatic Lockout Screen | Very Low | DOT does not have any plan for Mobile Devices' automatic lockout screen. |
| | Low | It is not DOT's priority to have an automatic lockout screen plan for Mobile Devices. |
| | Medium | It is important for DOT to have an automatic Lockout Screen plan for Mobile Devices but they are not very strict on that. |
| | High | DOT's Mobile Devices policies usually prohibit the circumvention of built-in security controls on mobile devices such as jailbreaking or rooting but they are not very strict on that. |
| | Very High | DOT's Mobile Devices policies always prohibit the circumvention of built-in security controls on mobile devices such as jailbreaking or rooting. |
| MAM | Very low | DOT does not provide any Mobile Application Management (MAM) plan for Mobile Devices' security. |
| | Low | DOT sometimes performs a Mobile Application Management (MAM) plan for Mobile Devices' security. |
| | Medium | It is not a priority for DOT to perform Mobile Application Management (MAM) for mobile devices' security maintenance, but it is part of their security plan. |
| | High | DOT usually performs Mobile Application Management (MAM) to maintain the |

|                        | Very High | DOT always performs or implements Mobile Application Management (MAM) to maintain the security of Mobile Devices and manages all changes to the devices' OS. |
|---|---|---|
| Remote Data Wipe | Very Low | DOT's IT department never has a plan for remote data wipe for Mobile Devices. |
|  | Low | DOT's IT department rarely provides remote data wipe or corporate data wipe for Mobile Devices. |
|  | Medium | DOT's IT department regularly provides remote data wipe or corporate data wipe for Mobile Devices. |
|  | High | DOT's IT department usually provides remote wipe or corporate data wipe for Mobile Devices. |
|  | Very High | DOT's IT department always provides remote wipe or corporate data wipe for all company-accepted BYOD devices or any mobile devices |
| Monitoring | Very Low | DOT does not have any controls on unauthorized software installation. |
|  | Low | DOT rarely controls unauthorized software installation. |
|  | Medium | DOT has a regular plan for unauthorized software installation, but not strictly. |
|  | High | DOT usually has controls in place to restrict and monitor the installation of unauthorized software onto the Mobile Devices. |
|  | Very High | DOT always has controls in place to restrict and monitor the installation of unauthorized software onto the Mobile Devices. |
| User name and Password | Very Low | DOT never asks employees to change the username and password of their Mobile Devices. |
|  | Low | DOT sometimes asks employees to change the username and password of their Mobile Devices but they are not required. |
|  | Medium | DOT regularly requires employees to change the username and password of their Mobile Devices. |
|  | High | DOT usually requires employees to change the username and password of their Mobile Devices. |
|  | Very High | DOT always strictly requires employees to change the username and password of their Mobile Devices. |

## 4. IMPLEMENTATION

As we mentioned in section 3.5 we implement only the Mobile Devices' factors and sub-factors as part of the MFL approach. Figure 6 shows the procedure of Mamdani FIS for Lost Devices Policies. Based on the MF derived from Group 1 and Group 2 we obtain the result of the output variable 'LostDevices' that is a sub-factor of MDM in a Triangular MF format. Figure 7 represents the MF plot for predefined Fuzzy sets in Group 1 that is limited in the range of 0 to 10 and determines each sub-factor's value derived from the security question. The value 10 indicates the highest security for the 'LostDevices' subfactor in Mobile Devices and 0 indicates the lowest one in DOT's network system.

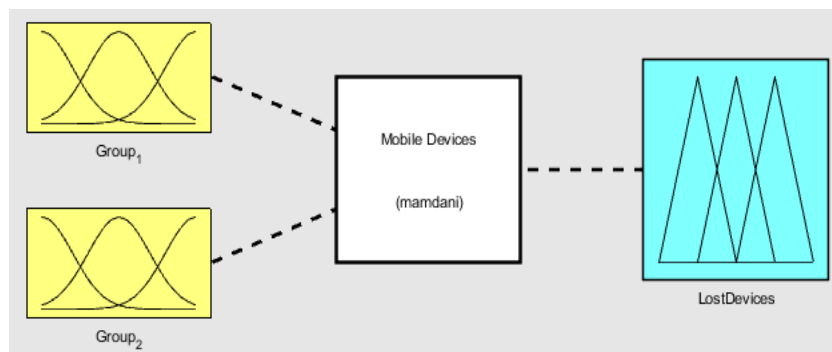

Figure 6. FIS for LostDevices.

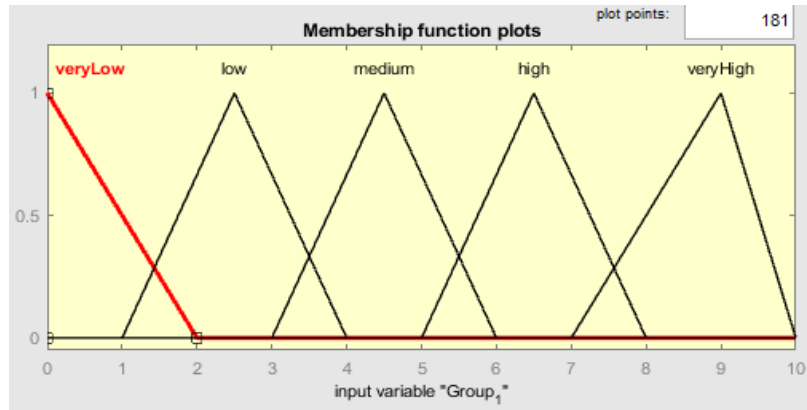

Figure 7. Fuzzy Subset and MF for Group 1.

In the FIS system, before the inference operations, the rules should be defined for the Fuzzy output variable. Thus, we apply `if-then` rules to map the input to the output. The main advantage of `if-then` rules is that they are evaluated in parallel, therefore, the order of rules does not matter. As we mentioned before, these rules are defined based on the expert's answers in the questionnaire that was derived from the DOT's network system. Thus, we define all terms we plan to use in the rules that interpret the values in the input vector and support to assign the appropriate values to the output vector. It is used for a combination of attributes based on the linguistic declaration.

Fuzzy rules are assigned to each group of subsets to provide a functional relationship between Fuzzy attributes. As shown in Figure 8 we defined 25 rules for each group in the Centroid (COA) model. The total number of rules depends on the number of subsets. In this case, we have two groups and each group has 5 subsets that result in 25 rules. All rules are assigned to the same weight 1, presented at the end of the line for each rule, however, they can be varied in the interval of [0,1].

```
11. If (Group_1 is medium) and (Group_2 is veryLow) then (LostDevices is low) (1)
12. If (Group_1 is medium) and (Group_2 is low) then (LostDevices is low) (1)
13. If (Group_1 is medium) and (Group_2 is medium) then (LostDevices is medium) (1)
14. If (Group_1 is medium) and (Group_2 is high) then (LostDevices is medium) (1)
15. If (Group_1 is medium) and (Group_2 is veryHigh) then (LostDevices is high) (1)
16. If (Group_1 is high) and (Group_2 is veryLow) then (LostDevices is low) (1)
17. If (Group_1 is high) and (Group_2 is low) then (LostDevices is medium) (1)
18. If (Group_1 is high) and (Group_2 is medium) then (LostDevices is medium) (1)
19. If (Group_1 is high) and (Group_2 is high) then (LostDevices is high) (1)
20. If (Group_1 is high) and (Group_2 is veryHigh) then (LostDevices is high) (1)
21. If (Group_1 is veryHigh) and (Group_2 is veryLow) then (LostDevices is medium) (1)
22. If (Group_1 is veryHigh) and (Group_2 is low) then (LostDevices is medium) (1)
23. If (Group_1 is veryHigh) and (Group_2 is medium) then (LostDevices is medium) (1)
24. If (Group_1 is veryHigh) and (Group_2 is high) then (LostDevices is high) (1)
25. If (Group_1 is veryHigh) and (Group_2 is veryHigh) then (LostDevices is veryHigh) (1)
```

Figure 8. Fuzzy Rules for LostDevices.

Each of the above `if-then` rules generates an output in the form of a Fuzzy set. To make a decision based on a single Fuzzy set, we need to apply the aggregation method (Figure 9) to combine all Fuzzy sets from `if-then` rules to a single Fuzzy set. The last plot in Figure 9 represents the aggregation of all combined Fuzzy sets. In this paper, we use the max-min technique for aggregation and the final value is achieved by the following equation:

```
Final Value= max(Group₁, Group₂,…, Groupₙ)
```

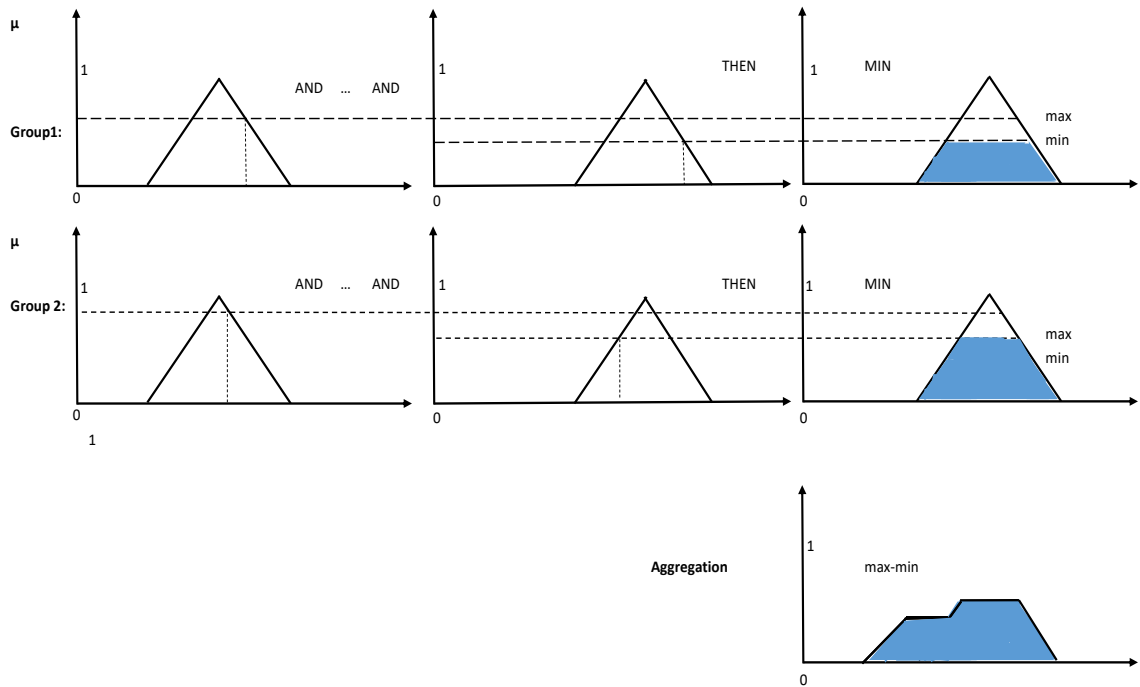

Figure 9. Aggregation in Fuzzy Sets.

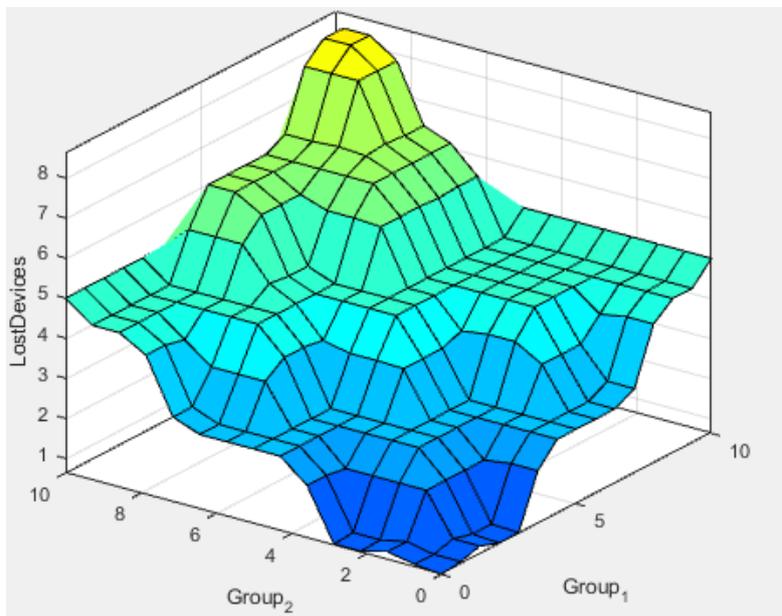

Figure 10. Output Curve for LostDevices.

The three-dimensional curve in Figure 10 depicts the mapping from Group1 and Group 2 to LostDevices. The vertical axis LostDevices represents the range of 0 to 9 that 0 indicates the least security arrangement (maximum vulnerability) for Mobile Devices and 9 indicates the maximum security plan (minimum vulnerability).

Fuzzy Inference processes are presented in Figure 11. It helps us to adjust the input values and obtain the corresponding aggregated output value for each Fuzzy rule. The first two columns of the plot depict the MF referenced by if-part of each rule. This part is called the antecedent and the last column indicates the MF referenced by then-part of each rule which is called the consequent. The average value for inputs and output is displayed on top of each column. Group 1 has an average value of 6.2 and Group 2 has an average value of 7.97 out of 10. The last column displays for LostDevice the result value 6.5 out of 10 (characterized by the last plot of the third column at the very right bottom) that indicates the weighted decision of the inference system based on the aggregation that depends on the input values. The value 6.5 indicates the security of that parameter in Mobile Devices. Therefore, the vulnerability of that is achieved by 10-6.5= 3.5 up to this point.

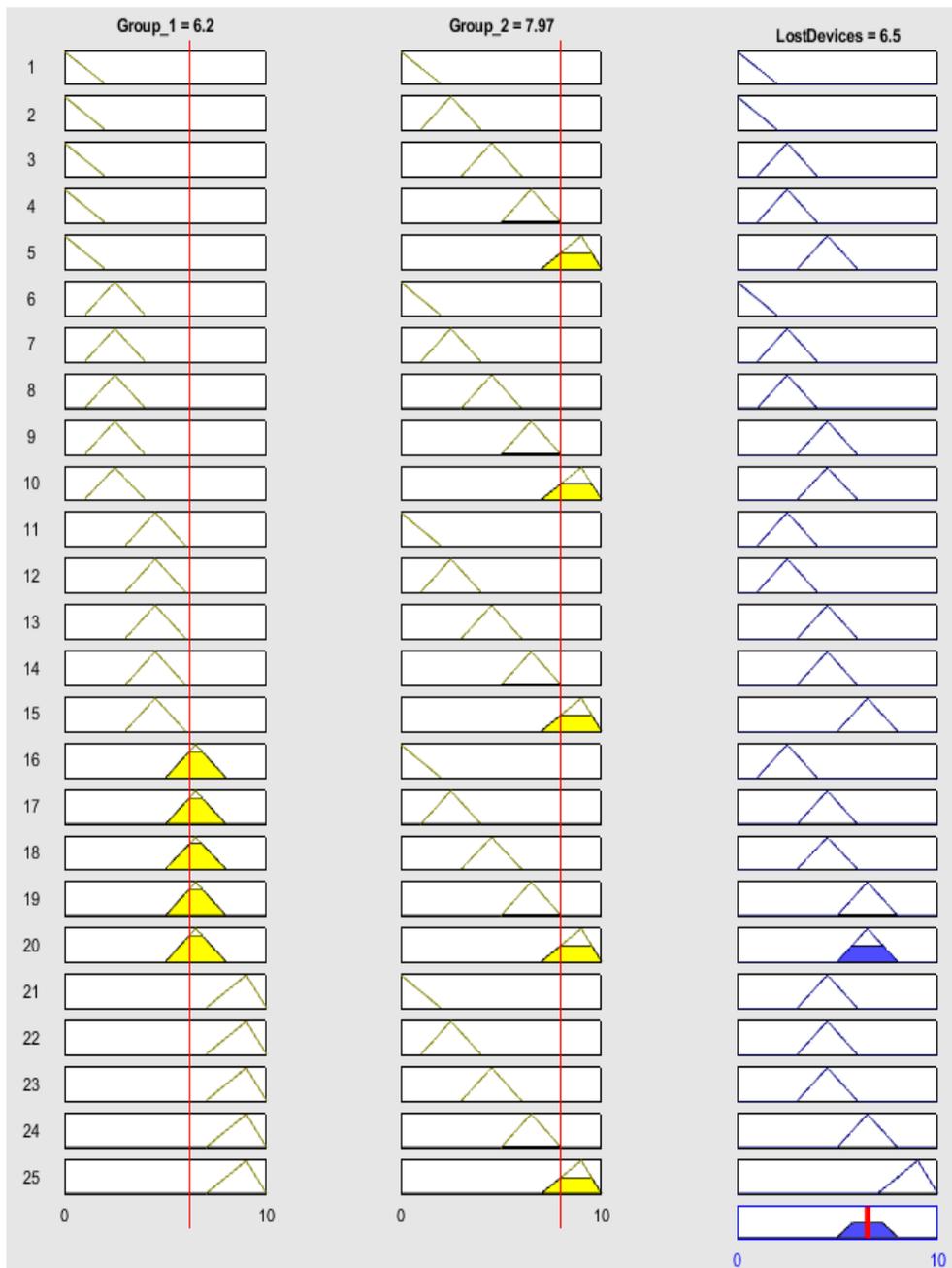

Figure 11. Fuzzy Inference Processes for LostDevices.

In order to achieve the vulnerability of the sub-factor Mobile Device Management System (MDM) as part of the factor Mobile Devices, we need to aggregate the output of the previous step's Fuzzy layer with the output of FIS for Group 3 and 4 in the next layer (Figure 5). As it displays in Figure 12 the output value for this Fuzzy layer is 4.58. Therefore, the output of these two layers is (6.5+4.58)/2=5.54 which represents the security level of Mobile Devices at this point. As a result, the vulnerability value for MDM is 4.46. This process will be continued until we obtain the final value for Mobile Devices as part of IoT in DOT.

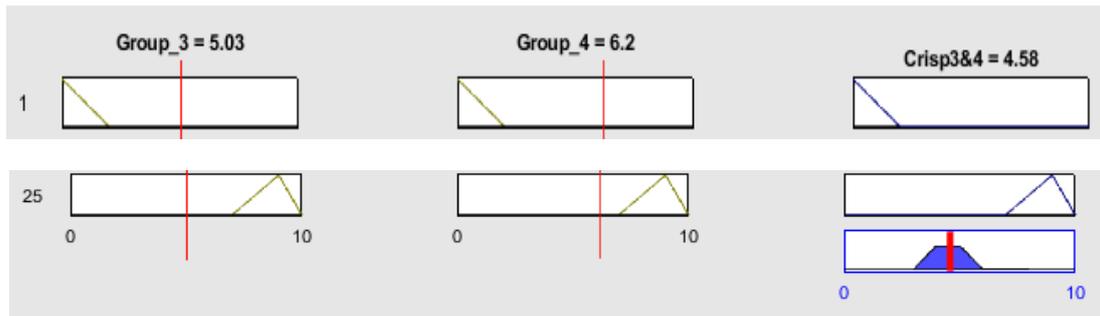

Figure 12. Fuzzy Inference Processes for Group 3 and 4.

## 5. CONCLUSIONS AND FUTURE WORK

As always network security experts describe the quality of security of a network in a human linguistic manner such as 'good' or 'relatively good', there would be a variety of interpretations for such description. Thus, our MFL methodology can evaluate quantitative vulnerability values using the Fuzzy Inference System. The findings of this study can be understood as a precise methodology of vulnerability analysis of network security, IoT, etc. in organizations. To our knowledge, this the first report of quantifying vulnerability based on security standards, GQM and without relying on other vulnerability measurement software such as CVSS and Nessus. Broadly translated our findings indicate that this approach can be applied not only in DOT but any agency that wants to measure the vulnerability of their network system quantitatively.

The future work of this research will be sending the questionnaire to the network security experts of DOT and measure the vulnerability of all security aspects of DOT such as physical security, Web Applications, Audit, etc. Moreover, future investigations are necessary to validate the kinds of conclusions that can be drawn from this study.

## REFERENCES

[1] B. Madan, K. Gogeva-Popstojanova, K. Vaidyanathan, and K. Trivedi, "Modeling and quantification of security attributes of software systems," *Proceedings International Conference on Dependable Systems and Networks, 2002*.

[2] M. U. A. Khan and M. Zulkernine, "Quantifying Security in Secure Software Development Phases," *2008 32nd Annual IEEE International Computer Software and Applications Conference*, 2008.

[3] P. Ralston, J. Graham, and J. Hieb, "Cyber security risk assessment for SCADA and DCS networks," *ISA Transactions*, vol. 46, no. 4, pp. 583–594, 2007.

[4] C. J. Alberts and A. J. Dorofee, *Managing Information Security Risks: The OCTAVESM Approach*. Addison-Wesley Professional, 2002.

[5] P.-C. Cheng, P. Rohatgi, C. Keser, P. A. Karger, G. M. Wagner, and A. S. Reninger, "Fuzzy Multi-Level Security: An Experiment on Quantified Risk-Adaptive Access Control," 2007 IEEE Symposium on Security and Privacy (SP 07), 2007.

[6] V. R. Basili and S. Green, "Software Process Evolution at the SEL," Foundations of Empirical Software Engineering, pp. 142–154.

[7] L. Zadeh, "Fuzzy sets," Information and Control, vol. 8, no. 3, pp. 338–353, 1965.

[8] J. Zhao and B. Bose, "Evaluation of membership functions for fuzzy logic controlled induction motor drive," *IEEE 2002 28th Annual Conference of the Industrial Electronics Society. IECON 02*.

[9] H.-J. Zimmermann, "Fuzzy Sets, Decision Making, and Expert Systems," 1987.

[10] W. Pedrycz, "Why triangular membership functions?," Fuzzy Sets and Systems, vol. 64, no. 1, pp. 21–30, 1994.

[11] N. Karnik, J. Mendel, and Q. Liang, "Type-2 fuzzy logic systems," IEEE Transactions on Fuzzy Systems, vol. 7, no. 6, pp. 643–658, 1999.

[12] W. V. Leekwijck and E. E. Kerre, "Defuzzification: criteria and classification," Fuzzy Sets and Systems, vol. 108, no. 2, pp. 159–178, 1999.

[13] E. Mamdani, "Advances in the linguistic synthesis of fuzzy controllers," International Journal of Man-Machine Studies, vol. 8, no. 6, pp. 669–678, 1976.

[14] E. Erturk and E. A. Sezer, "Software fault prediction using Mamdani type fuzzy inference system," International Journal of Data Analysis Techniques and Strategies, vol. 8, no. 1, p. 14, 2016.

[15] A. S. Sodiya, S. A. Onashoga, and B. Oladunjoye, "Threat Modeling Using Fuzzy Logic Paradigm," *Proceedings of the 2007 InSITE Conference*, 2007.

[16] P. Kamongi, S. Kotikela, M. Gomathisankaran, and K. Kavi, "A methodology for ranking cloud system vulnerabilities," *2013 Fourth International Conference on Computing, Communications and Networking Technologies (ICCCNT)*, 2013.

[17] I. V. Anikin, "Information security risk assessment and management method in computer networks," *2015 International Siberian Conference on Control and Communications (SIBCON)*, 2015.

[18] "Common Vulnerability Scoring System SIG," *FIRST*. [Online]. Available: https://www.first.org/cvss/. [Accessed: 03-Feb-2020].

[19] G. Kuang, X. Wang, and L. Yin, "A fuzzy forecast method for network security situation based on Markov," *2012 International Conference on Computer Science and Information Processing (CSIP)*, 2012.

[20] KDD Cup 1999 Data. [Online]. Available: http://kdd.ics.uci.edu/databases/kddcup99/kddcup99.html.

[21] "2000 DARPA Intrusion Detection Scenario Specific Datasets," MIT Lincoln Laboratory. [Online]. Available: https://www.ll.mit.edu/r-d/datasets/2000-darpa-intrusion-detection-scenario-specific-datasets.

[22] A. Broring, S. Schmid, C.-K. Schindhelm, A. Khelil, S. Kabisch, D. Kramer, D. L. Phuoc, J. Mitic, D. Anicic, and E. Teniente, "Enabling IoT Ecosystems through Platform Interoperability," *IEEE Software*, vol. 34, no. 1, pp. 54–61, 2017.

[23] R. Williams, E. Mcmahon, S. Samtani, M. Patton, and H. Chen, "Identifying vulnerabilities of consumer Internet of Things (IoT) devices: A scalable approach," *2017 IEEE International Conference on Intelligence and Security Informatics (ISI)*, 2017.

[24] "Introducing Nessus," *Nessus Network Auditing*, pp. 27–43, 2004.

[25] M. Patton, E. Gross, R. Chinn, S. Forbis, L. Walker, and H. Chen, "Uninvited Connections: A Study of Vulnerable Devices on the Internet of Things (IoT)," *2014 IEEE Joint Intelligence and Security Informatics Conference*, 2014.

[26] "Leading the IoT - gartner.com." https://www.gartner.com/imagesrv/books/iot/iotEbook_digital.pdf.

[27] Boehm, B. W. & Papaccio, P. N. "Understanding and Controlling Software Costs." IEEE Transactions on Software Engineering SE-4, 10 (October 1988): 1462-77.

[28] Ahl V (2005) An experimental comparison of five prioritization methods. Master's Thesis, School of Engineering, Blekinge Institute of Technology, Ronneby, Sweden.

[29] V. R. Basili and S. Green, "Software Process Evolution at the SEL," Foundations of Empirical Software Engineering, pp. 142–154.

[30] M. Shepperd, "Practical software metrics for project management and process. improvement," Information and Software Technology, vol. 35, no. 11-12, p. 701, 1993.

[31] M. Shepperd, "Practical software metrics for project management and process improvement," Information and Software Technology, vol. 35, no. 11-12, p. 701, 1993.

[32] M. Shojaeshafiei, L. Etzkorn, and M. Anderson, "Cybersecurity Framework Requirements to Quantify Vulnerabilities Based on GQM," SpringerLink, 04-Jun-2019. [Online]. Available: https://link.springer.com/chapter/10.1007/978-3-030-31239-8_20.

[33] "The Iso 27001 Risk Assessment," Information Security Risk Management for ISO 27001/ISO 27002, third edition, pp. 87–93, 2019.

[34] H. Susanto and M. N. Almunawar, "Information Security Management Systems," 2018.

[35] J. T. Force, "Security and Privacy Controls for Information Systems and Organizations," CSRC, 15-Aug-2017. [Online]. Available: https://csrc.nist.gov/publications/detail/sp/800-53/rev-5/draft.

[36] L. Pokoradi and B. Szamosi, "Fuzzy failure modes and effects analysis with summarized center of gravity defuzzification," *2015 16th IEEE International Symposium on Computational Intelligence and Informatics (CINTI)*, 2015.

[37] E. Ngai and F. Wat, "Fuzzy decision support system for risk analysis in e-commerce development," Decision Support Systems, vol. 40, no. 2, pp. 235–255, 2005.

[38] J. Kacprzyk, "Group decision making with a fuzzy linguistic majority," Fuzzy Sets and Systems, vol. 18, no. 2, pp. 105–118, 1986.

[39] A. Lotfi and A. Tsoi, "Learning fuzzy inference systems using an adaptive membership function scheme," *IEEE Transactions on Systems, Man and Cybernetics, Part B (Cybernetics)*, vol. 26, no. 2, pp. 326–331, 1996.